\documentstyle[12pt]{article}
\title{Vacuum energy:  quantum hydrodynamics vs quantum gravity}
\author{G.E. Volovik\\
Low Temperature Laboratory, 
Helsinki University of Technology\\
P.O.Box 2200, FIN-02015 HUT, Finland\\
and\\
L.D. Landau Institute for Theoretical Physics, 
 Moscow\\
}
\begin{document}
\maketitle
\begin{abstract}
{ We compare quantum hydrodynamics and quantum gravity. They
share many common features. In particular, both have quadratic
divergences, and both lead to the problem of the vacuum energy,
which in the quantum gravity transforms to the cosmological
constant problem. We show that in quantum liquids the vacuum
energy density is not determined by the quantum
zero-point energy of the phonon modes. The energy density
of the vacuum is much smaller and is determined by the
classical macroscopic parameters of the liquid including the
radius of the liquid droplet.  In the same manner the
cosmological constant is not determined by the 
zero-point energy of quantum fields. It is much
smaller and is determined by the classical macroscopic
parameters of the Universe dynamics:  the Hubble radius, the
Newton constant  and the energy density of matter.
The same may
hold for the Higgs mass problem:  the
quadratically divergent
quantum correction to the Higgs potential mass term is also
cancelled by the microscopic (trans-Planckian) degrees of
freedom due to thermodynamic stability of the whole quantum vacuum.
 } 
\end{abstract}


\section{Introduction.} 

The problem of quantum hydrodynamics is at least
65 years old (see quantization of the macroscopic dynamics of
liquid in the first Landau paper on superfluidity of $^4$He
\cite{Landau41}). It is almost as old as the problem of quantum
gravity
\cite{Stachel}. Quantum hydrodynamics and quantum gravity share
many common features (e.g. both have quadratic divergences) and
probably they will have the common destiny. The main message
from quantum hydrodynamics to quantum gravity is that most
probable the quantum gravity cannot be constructed, because
quantum hydrodynamics cannot be constructed.

Of course, one can quantize sound waves in hydrodynamics to
obtain quanta of sound waves --  phonons. Similarly one can
quantize gravitational waves in general relativity to obtain
gravitons. But one should not use the low-energy quantization for
calculation of the radiative corrections which contain Feynman
diagrams with integration over high momenta. In particular, the
effective field theory  is not appropriate for the calculation
of the vacuum energy in terms of the zero-point energy of quantum
fields. The latter leads to the cosmological
constant problem in gravity
\cite{Weinberg,Padmanabhan}, and to the similar paradox for the
vacuum energy in quantum hydrodynamics: in both cases the
vacuum energy estimated using the effective theory is by many
orders of magnitude too big.  We know how this paradox is
solved in quantum liquids, and we may expect that the same
general arguments based on the thermodynamic stability of the
ground state of the quantum liquid are applicable to the quantum
vacuum.

There is another big discrepancy between theory and experiment,
which is called the hierarchy problem in the Standard Model 
\cite{HiggsMassProblem}. It is believed that
 the mass of the Higgs boson is on the order of or somewhat larger
than the mass of the gauge boson:
$m_{\rm H}^2
\sim M^2_Z$. For example, in analogy with the effect of Cooper pairing in
superconductivity, $m_{\rm H}$ can
be equal to $2m_t$, where $m_t$ is the mass for the top quark
\cite{VeltmanBook,NambuBCS}. However the radiative correction to the Higgs mass
is quadratically diverging, and is determined by the ultraviolet cut-off.
If one chooses the natural GUT or Planck scale as the cutoff
energy, one obtains that the Higgs boson mass must be extremely
large: $m_{\rm H}^2 \sim \pm 10^{26}M^2_Z$ and  $m_{\rm
H}^2
\sim \pm 10^{34}M^2_Z$ correspondingly (the sign of the radiative  correction is determined by
the relevant fermionic and bosonic content of the theory and
the cut-off scheme). We argue, that this discrepancy is related
to the problem of the vacuum energy, and thus the same
thermodynamic arguments which have been used for the cosmological
constant problem can be applicable to the hierarchy problem.

\section{Classical hydrodynamics of quantum liquid} 

Let us consider the hydrodynamics of the isotropic superfluid  
liquid at $T=0$ (such as bosonic superfluid $^4$He  and
fermionic superfluid $^3$He-B \cite{VollhardtWolfle}). Though
the superfluid liquid is essentially quantum, its macroscopic
low-frequency dynamics is classical and is represented by
classical hydrodynamics. It is background independent, i.e. it does not depend on details of the
underlying microscopic physics, and is the same for fermionic
and bosonic liquids.  The equations of the non-relativistic
superfluid hydrodynamics (in the absence of quantized vortices)
are the Hamilton equations for the canonically conjugated fields:
\begin{equation}
\partial_t\rho=\frac{\delta H}{\delta
\phi}~~,~~\partial_t\phi=-\frac{\delta H}{\delta \rho}
 ~. 
\label{HamiltonEqs}
\end{equation}
 Here $\rho$ the mass density;   $\phi$ is the
velocity potential (in the absence of quantized vortices
the superfluid velocity is potential:  ${\bf v}=\nabla\phi$);
the Hamiltonian is the energy functional of the liquid expressed
in terms of
 $\rho$ and   $\phi$:
\begin{equation}
H=\int d^3r \left[ \frac{1}{2}\rho (\nabla\phi)^2
+\tilde\epsilon(\rho)\right]
 ~, 
\label{ClassicalEnergy}
\end{equation}
where
$\tilde\epsilon(\rho)=\epsilon(\rho)-\mu\rho$;
$\epsilon(\rho)$  is the energy density of the liquid expressed
in terms of the liquid density; and $\mu$ is chemical potential
-- the Lagrange multiplier which takes into account the mass
conservation $\int d^3r \rho=Const$. At fixed chemical potential
$\mu$, this functional has minimum at ${\bf v}=0$ and 
$\rho=\rho_0(\mu)$, where the equilibrium density $\rho_0$ is
determined by equation $d\tilde\epsilon/d\rho=0$ (or
$d\epsilon/d\rho=\mu$). This is the ground state of the liquid,
with the relevant thermodynamic potential -- the energy density $\tilde\epsilon(\rho_0)$ -- and the total relevant energy
\begin{equation}
E_0=V\tilde\epsilon(\rho_0) 
 ~. 
\label{ClassicalEnergy}
\end{equation}
Note again that this consideration is completely classical and
operates with quantities  $\rho_0$ and $\tilde\epsilon(\rho_0)$,
which are the classical output of the quantum system: the
superfluid $^4$He and superfluid $^3$He-B are  systems of
strongly correlated, strongly interacting and highly entangled
helium atoms governed by the laws of quantum mechanics. The
reason for classicality is the macroscopic character of the
collective motion.

\section{Quantized hydrodynamics} 

What happens if we try to construct the quantum
hydrodynamics, i.e. to quantize the hydrodynamic motion of
the liquid determined by Eqs.(\ref{HamiltonEqs})? Using
hydrodynamic variables only we are unable to reconstruct the
whole microscopic Hamiltonian for the interacting atoms. This is
because from the big realm of the complicated quantum motion of
the $^4$He atoms we have chosen only the hydrodynamic modes
whose  wavelengths are much bigger than the inter-atomic spacing
$a$, which plays the role of the Planck length:  $ka\ll 1$.
That is why, what we can do at best is to quantize the sound
modes. But even in this case there is a danger of the double
counting,  because starting from the quantum
system we have obtained the classical behavior of soft variables
$\rho$ and $\phi$, and now we are trying to quantize them again.
In particular, the energy
$E_0$ in Eq.(\ref{ClassicalEnergy}) is the whole energy of the
quantum liquid, and it already includes from the very
beginning the energy of those  degrees of freedom which
are described in terms of phonons. Let us see how this double
counting typically occurs.

The conventional quantization procedure for sound waves in the background
of the state with $\rho=\rho_0$ and $\phi=0$ is the
introduction of the commutation relations for the canonically
conjugated variables:
\begin{equation}
\left[ \hat{\phi}_{\bf k},\hat{\rho}_{{\bf
k}'}\right]=i\hbar \delta_ {{\bf k} {\bf k}'}~~,
\label{commutator}
\end{equation}
where
\begin{equation}
\hat\rho({\bf r})=\rho_0 +\frac{1}{\sqrt{V}}\sum_{\bf k}
\left( \hat\rho_{\bf k}e^{i{\bf k}\cdot{\bf r}} + c.c. \right)
 ~, 
\label{rho}
\end{equation}
\begin{equation}
\hat{\phi}({\bf r})= 
 \frac{1}{\sqrt{V}}\sum_{\bf k}
\left( \hat{\phi}_{\bf k}e^{i{\bf k}\cdot{\bf r}} + c.c. \right)
 ~, 
\label{v}
\end{equation}
Introducing these quantum fluctuations to Eq.
(\ref{ClassicalEnergy}) one obtains the quantum Hamiltonian as
the sum of the ground state energy and the Hamiltonians for
quantum oscillators:
\begin{equation}
\hat H=E_0+ \frac{1}{2}\sum_{\bf k}
\left(\rho_0 k^2|\hat\phi_{\bf k}|^2  + 
\frac{c^2}{\rho_0}|\hat\rho_{\bf k}|^2 
\right)=
E_0+  \frac{1}{2} \sum_{\bf k}   \hbar \omega_{\bf k}
\left(a_{\bf k} a_{\bf k}^\dagger  +  a_{\bf k}^\dagger a_{\bf
k}\right)
 ~. 
\label{Oscillators}
\end{equation}
where $\omega_{\bf k}=ck$; and the speed of sound $c$ is given by
$ c^2=\rho_0 d^2\epsilon/d\rho^2|_{\rho_0}$.

There is nothing bad with this
quantization, if we are constrained by condition $ka\ll 1$.
However, if we start to apply this consideration to the
diverging quantities such as the vacuum energy, we are in
trouble.

\section{Vacuum energy in quantum hydrodynamics} 

The vacuum energy -- the ground state of the Hamiltonian
(\ref{Oscillators}) --  contains the zero-point energies of
quantum oscillators:
\begin{equation}
\left<vac\left|\hat H\right| vac\right>=E_0+ \frac{1}{2}\sum_{\bf
k}
  \hbar \omega_{\bf k}~~. 
\label{ZPEnergy}
\end{equation}
However, the vacuum expectation value of $\hat H$
must be equal to  $E_0$  by definition. This is because $E_0$ is not
the ``bare'' energy, but is the total relevant energy of the liquid, which includes
all quantum degrees of freedom of the liquid.
Thus the naive application of the zero-point energy leads to the
paradoxical conclusion that
\begin{equation}
 \frac{1}{2}\sum_{\bf k}   \hbar \omega_{\bf k}=0~~. 
\label{ZeroZPE}
\end{equation}
This contradicts to our intuition that the zero-point fluctuations
give for the vacuum energy the estimate $ \tilde\epsilon_{\rm
zp}\sim 
\hbar ck_{\rm uv}^4$, where
$k_{\rm uv}\sim 1/a$ is the ultraviolet cut-off. However, the 
equation (\ref{ZeroZPE}) simply means that the zero-point energy
of phonons had already been included into
the original $E_0$ together with all other modes, i.e. by writing Eq.(\ref{ZeroZPE}) we simply prevent
 the double counting for the vacuum energy. 
Thus the correct form of the Hamiltonian for phonons
must be
\begin{equation}
\hat H=E_0+    \sum_{\bf k}   \hbar \omega_{\bf k}
 a_{\bf k}^\dagger a_{\bf
k}
 ~. 
\label{OscillatorsCorrect}
\end{equation}

One may ask what is the role of the quantum fluctuations of the hydrodynamic field.
 Do they provide the main or substantial part of $E_0$?
To see this
let us consider the vacuum energy of superfluid
liquid in case when an external pressure $P$ is applied to the
liquid. According to the Gibbs-Duhem relation which is valid for
the equilibrium states one has:
\begin{equation}
P=TS+ \mu\rho_0 -\epsilon(\rho_0)~. 
\label{GibbsDuhem}
\end{equation}
At $T=0$ one obtains that the relevant vacuum energy density in Eq. (\ref{ClassicalEnergy})  is
regulated by external pressure
\begin{equation}
\tilde\epsilon=-P~. 
\label{GibbsDuhem2}
\end{equation}
For the positive external pressure $P>0$, one obtains the
negative  energy density of the 
vacuum, $\tilde\epsilon<0$, which  certainly cannot be
obtained  by summation of the positive zero-point energies of
phonons.  Thus the contribution of zero-point energies of phonons to $E_0$ 
gives no idea on the total value of  $E_0$, since it
may even give the wrong sign of the vacuum energy. 

Furthermore, for the  superfluid
helium liquid isolated from the environment, the pressure $P=0$,
and thus the vacuum energy density and vacuum energy are zero:
$\tilde\epsilon(\rho_0)=0$, $E_0=0$. For the finite system --  the helium droplet --
 the vacuum  energy density
becomes  non-zero due to the capillary pressure:
\begin{equation}
\tilde\epsilon=-P=-\frac{2\sigma}{R}~. 
\label{CapillaryPr}
\end{equation}
   It is expressed through 
the classical parameters of the liquid droplet: its radius $R$ and 
surface tension $\sigma$.
 When we compare this physical
result with the naive estimation which only takes into account the
 contribution of the phonon zero-point energy 
$\tilde\epsilon_{\rm zp}\sim    \hbar ck_{\rm uv}^4$,  one finds
that  their ratio is determined by the ratio of
the quantum microscopic to the classical macroscopic scales:
$\tilde\epsilon_{\rm true}/\tilde\epsilon_{\rm zp}\sim a/R$.
For the macroscopic bodies the discrepancy is big.

\section{Application to quantum vacuum}

This demonstrates that the vacuum energy is determined by the
macroscopic thermodynamic laws and is not related to the
diverging contribution of the zero-point motion of phonons. The
result
$E_0=0$ for the self-sustained homogeneous systems shows that in this
system the large positive contribution $\tilde\epsilon_{\rm
zp}\sim   \hbar ck_{\rm uv}^4$ of phonons is completely compensated 
without any fine-tuning by the
other quantum degrees of freedom, i.e. by the microscopic
(atomic $\equiv$ trans-Planckian) degrees of freedom which 
cannot be described in terms of the effective (hydrodynamic) field  
\cite{Book}.
 
This compensation which occurs in the homogeneous vacuum
 does not prohibit the Casimir effect in the systems with boundaries. If the energy
difference between two vacua comes solely from the
long-wavelength physics, it is within responsibility of the
phonon (photon) modes and can be calculated using their
zero-point energies.

One can immediately apply this lesson to the quantum gravity: the
vacuum energy density $\epsilon_{\rm vac}$  (or cosmological
constant)  is not renormalized by
the zero-point energies of quantum fluctuations of the low-energy
modes. It is meaningless to represent the vacuum
energy as the sum over zero-point oscillations
$\frac{1}{2}\sum_{\bf k}  \hbar \omega_{\bf k}$,  and to
estimate the vacuum energy density as
$\epsilon_{\rm zp}\sim \hbar ck_{\rm uv}^4$. The vacuum energy
(and thus the cosmological constant) is the final classical
output of the whole quantum vacuum with all its degrees of
freedom, sub-Planckian and trans-Planckian. It is regulated by
macroscopic physics, and obeys the macroscopic thermodynamic
laws.  The thermodynamic Gibbs-Duhem relation (analog of
Eq.(\ref{GibbsDuhem2})) must be satisfied for the equilibrium vacuum,
and it does follow from the cosmological term in  Einstein action:
\begin{equation}
\Lambda=\epsilon_{\rm vac}
=-P_{\rm vac} ~. 
\label{GibbsDuhemEinst}
\end{equation}
For the vacuum isolated from the
"environment" (free vacuum), the pressure is zero  and thus the vacuum energy
density is  zero too, $\epsilon_{\rm vac}=-P_{\rm vac}=0$
\cite{Book,AnnPhys}. This means that the natural value of the
cosmological constant in the equilibrium homogeneous time-independent 
free vacuum
is
$\Lambda=0$  rather than
 the contribution $\Lambda \sim \hbar ck_{\rm uv}^4$ of the zero point energies
 of the effective quantum fields which is completely compensated. 

In the case of the developing Universe polluted by matter, 
the vacuum energy is disturbed, and the compensation is not complete. But again the
natural value of
$\Lambda$ is determined not by the quantum zero-point energy, but
(as in quantum liquid in Eq.(\ref{CapillaryPr})) by the classical
macroscopic parameters of the Universe dynamics:  the Hubble
radius $R$ of the Universe, the Newton constant $G$ and the
energy density of matter $\rho_{\rm mat}$. This implies that,
depending on the details of the process,  one has $\Lambda\sim
\rho_{\rm mat}$, or
$\Lambda\sim 1/GR^2$ (or $\Lambda$ is given by some combination
of these factors). Both estimates are comparable to the measured
value of
$\Lambda$, and are much smaller than the naive estimation of
the zero-point energy of quantum fields: $\Lambda_{\rm
true}/\Lambda_{\rm zp}
\sim a^2/R^2 \sim 10^{-120}$.  As in quantum hydrodynamics
this contains the ratio of the quantum microscopic scale $a=1/k_{\rm uv}=\hbar
c/E_{\rm Planck}$ to the classical macroscopic scale $R$. 

\section{Higgs mass problem}

As distinct from quantum gravity,  the Standard Model of
electroweak and strong  interactions is the low-energy
effective theory which can exist in the quantum form. The
condensed matter experience demonstrates that this effective
theory can be emergent. In condensed matter systems with point
nodes in the energy spectrum of fermions,  the effective gauge
fields and chiral fermions gradually emerge  
 in the low-energy corner. The point nodes are protected by
topology in momentum space and thus are generic \cite{Book},
that is why the effective theory of the Standard-Model type is
generic. Though this effective theory is the final output  of the
underlying quantum system, it can be quantized again, in spite
of the ultraviolet divergences. The reasons for that is that
above the symmetry breaking electroweak scale, i.e. at $ k^2\gg
M_Z^2$, the divergences are logarithmic, of the type
$\ln (k_{\rm}^2/M_Z^2)$. Since the logarithm is concentrated
mostly in the sub-Planckian  region $k_{\rm uv}\gg k^2 \gg
M_Z^2$, it is within the jurisdiction of the low-energy
effective theory.

 At $k^2 \sim M_Z^2\sim 10^{-34}E_{\rm Planck}^2$ the symmetry
breaking occurs and the non-zero vev of the Higgs field
develops.  One may expect that  at these extremely low energies
as compared to the Planck or GUT scale, the effective theory
must work well. Instead  the ultraviolet problem arises. In
particular, if one tries to calculate the quantum radiative
corrections to the mass of the Higgs boson, the quadratic
divergence occurs, which is outside the jurisdiction of the
Standard Model. What can one say on this problem using the
condensed matter experience?

To describe the electroweak transition,  the action for the gauge fields and quarks and leptons 
is supplemented by the action for the Higgs field 
\begin{equation}
S_{\rm Higgs}=  \int d^4x \left(\frac{1}{4}  \lambda
(\phi^2 -\phi_0^2)^2 + (c^2\nabla \phi)^2 -(\partial_t \phi)^2
\right)
 ~,
\label{HiggsLagr}
\end{equation}
and the  terms describing interaction of the Higgs field with fermions and gauge fields.
The Higgs field $\phi$ here is a weak doublet.  The Hamilton
function for the Higgs field is:
\begin{equation}
H_{\rm Higgs}=  E_0 + \int d^3x \left(\frac{1}{4} 
\lambda (\phi^2 -\phi_0^2)^2 + (c^2\nabla \phi)^2 +(\partial_t
\phi)^2
\right)
 ~. 
\label{HiggsHam}
\end{equation}
The Higgs field is massive with mass
\begin{equation}
m^2_{\rm H}=\lambda \phi_0^2~~
. 
\label{MasseHiggs}
\end{equation}
From the  condensed matter point of view,  the mass $m^2_{\rm
H}$ and the vacuum energy $E_0$ are the final classical output of
the whole underlying quantum vacuum. We know that the
energy of the whole vacuum must be zero according to the
thermodynamic stability of the free vacuum: 
$E_0=H(\phi=\phi_0)=0$.   If one does not take into account the
possible infrared  anomalies, the equation (\ref{HiggsHam}) with
$E_0=0$ is the general form satisfying the stability condition at
$T=0$ in the absence of the external environment. In the
effective theory, the information from the underlying physics is
thus lost, and the type of the effective theory only depends on
the symmetry and topology of the system. 

Two parameters of the
effective theory -- $\lambda$ and the equilibrium value
$\phi_0$  of the Higgs field -- can be considered as
phenomenological. Though these parameters are determined by the
microscopic (Planck) physics, their values may essentially
differ from the microscopic scales. Examples are provided by
superconductivity of metals and superfluidity of Fermi liquids,
whose energy scale  is exponentially small compared to the
corresponding microscopic energy  scale $E_F$ (the Fermi
energy): $\lambda\phi_0^2\sim E_F^2\exp(-1/g) $.  Since the
effective coupling $g$ is  typically small,  $g\ll 1$, this 
leads to the macroscopic energy and length scales for the
effective Ginzburg-Landau theory of superconductivity. In
Rhodium metal, which has the lowest  transition temperature
observed in metals, $T_c\sim 0.3$mK \cite{Knuuttila}, one has
$\lambda\phi_0^2\sim 10^{-14}E_F^2$. In the fermionic superfluid
liquid $^3$He, the coupling $g$ is small due to the many-body
effects, and one has $\lambda\phi_0^2\sim 10^{-6}E_F^2$; the
superfluidity of the $^3$He atoms in dilute $^3$He-$^4$He
mixture has not yet been found, which suggests that  in this
system $\lambda\phi_0^2< 10^{-8}E_F^2$.

Let us see what happens if we take into account the
zero-point energy of quantum fields, including the zero-point
energy of the Higgs field 
$\phi$.  In condensed mater this means that  we quantize the
system again, but now instead of the whole system of atoms we 
are only dealing with the collective bosonic and fermionic
fields which enter the effective theory.

Introducing the quantum operators for the Higgs field
\begin{equation}
\hat\phi=\phi_0 +\frac{1}{\sqrt{V}}\sum_{\bf k}
\left( \hat\phi_{\bf k}e^{i{\bf k}\cdot{\bf r}} + c.c. \right)
 ~, 
\label{expansion}
\end{equation}
one obtains the Hamiltonian for the Higgs bosons:
\begin{equation}
\hat H_{\rm Higgs}= E_{0}+  \frac{1}{2} \sum_{\bf
k}   \hbar
\omega_{\bf k}
\left(a_{\bf k} a_{\bf k}^\dagger  +  a_{\bf k}^\dagger a_{\bf
k}\right)
 ~. 
\label{OscillatorsHiggs}
\end{equation}
The vacuum energy contains now the zero-point energy
of $\phi$-field, and for completeness one must also add the
zero-point energy of other bosonic fields of the Standard Model 
-- the gauge fields -- and the negative energy of the Dirac
vacuum of fermions (quarks and leptons):
\begin{equation}
E_{\rm zp}=\frac{1}{2}\sum_{{\bf k}b}
  \hbar \omega_{{\bf k}b}-\sum_{{\bf k}f}
  \hbar \omega_{{\bf k}f}~~. 
\label{ZPEnergy}
\end{equation}
From the condensed matter experience, it follows that we must
forget about the zero-point energy of the effective fields. They
have already been included into the vacuum energy $E_0$ together
with the energies of microscopic degrees of freedom. Moreover,
the total vacuum energy must be zero in equilibrium, which means
that the trans-Planckian physics fully compensates the
zero-point energy contribution of the sub-Planckian modes
without any fine-tuning. 

Let us suppose that we are not aware on this
thermodynamic principle of perfect compensation. Then we must
seriously consider the zero-point energy of bosonic and fermionic quantum 
fields (massless and massive) and estimate its contribution to radiative correction
to the Higgs mass.  Due to interaction of
the Higgs field with gauge bosons and fermions, all the fermions
and also $W$- and $Z$-bosons become massive:
\begin{equation}
\omega_{{\bf k}f}^2 =m_f^2+c^2k^2 ~~,~~\omega_{{\bf k}b}^2
=m_b^2+c^2k^2
 ~,  
\label{Spectrum}
\end{equation}
and masses of fermions
and bosons depend on the Higgs field
\begin{equation}
m^2_{f,b}=\lambda_{f,b} \phi_0^2~~
,  
\label{Masses}
\end{equation}
(for the Higgs field itself the corresponding $\lambda_{\rm H}=
\lambda$ in eq.(\ref{MasseHiggs}); while the photon remains massless, $\lambda_A=0$).
This dependence leads to
the $\phi_0^2$ term in the zero-point energy, which must be
identified with  the additional mass term for the Higgs field coming
from the quantum effects: 
\begin{equation}
\delta (m^2_{\rm H})= \frac{1}{2}\frac{d^2E_{\rm
zp}}{d\phi_0^2}\approx  \frac{1}{2}\sum_{{\bf k}b}
  \frac{\lambda_b}{\omega_{{\bf k}b}}-\sum_{{\bf k}f}
   \frac{\lambda_f}{ \omega_{{\bf k}f}}~. 
\label{RadCorr1}
\end{equation}
The sum in Eq.(\ref{RadCorr1}) diverges quadratically.
Neglecting fermion masses except for that of the heaviest fermion
-- the top quark with mass $m_t$ --  one obtains  for
 the radiative correction to the Higgs mass:
\begin{equation}
\delta (m^2_{\rm H})\sim \lambda k_{\rm uv}^2  \frac{m^2_{\rm H}
+2M_Z^2+ 4M_W^2 - 12m_t^2}{m^2_{\rm H}}  ~~. 
\label{RadCorr2}
\end{equation}
With $M_W\sim M_Z \sim  m_t \sim m_{\rm H}\sim
10^{2}-10^{3}$GeV, $\lambda\sim 1$ and $k_{\rm uv}\sim
10^{16}-10^{19}$GeV, the quantum correction $\delta (m^2_{\rm
H})$ is by many orders of magnitude bigger than $m^2_{\rm H}$
itself;   this is the hierarchy problem in the Standard Model
(see review paper \cite{HiggsMassProblem}). From the
condensed-matter point of view, the origin of this huge
discrepancy is the same as in the cosmological constant problem:
it is the diverging zero-point energy of quantum fields. The
general form of the zero point energy in Eq.(\ref{ZPEnergy}) is
\begin{equation}
E_{\rm zp}= a_4k_{\rm uv}^4 +a_2 k_{\rm uv}^2 m^2_{\rm H} +
a_0
 m^4_{\rm H} \ln (k_{\rm uv}^2/m^2_{\rm H})~~, 
\label{ZPEnergy2}
\end{equation}
where $|a_4|$, $|a_2|$ and $|a_0|$ are of order
unity.  The main contribution of zero-point energy to the
vacuum energy and cosmological constant diverges quartically,
while its contribution to the Higgs mass diverges quadratically.
The term with
$a_4$ gives too large value for the vacuum energy, which in
systems with gravity transforms to the large cosmological
constant
$\Lambda\sim k_{\rm uv}^4$ leading to the main cosmological
constant problem \cite{Weinberg,Padmanabhan}. The term with
$a_2$ leads to the hierarchy problem in the Standard Model, since
 it gives $\delta (m^2_{\rm H})=(1/2) d^2E_{\rm
zp}/d\phi_0^2\sim k_{\rm uv}^2$ according to Eq.(\ref{MasseHiggs}).
Both problems come from the same
estimation of the vacuum energy as the zero point energy.
 
The exact calculation and also the thermodynamic
analysis demonstrate that in equilibrium the total vacuum energy
is zero, $E_{\rm vac}=0$, if the system is isolated from the
environment. The zero-point energy of quantum fields, which is
only a part of the whole energy of the quantum vacuum, is  thus
fully compensated by the microscopic degrees of freedom. This
suggests the solution to the cosmological constant problem, and
also makes doubtful the use of the zero-point energy for the
estimation of the mass of Higgs boson, since the quadratically
diverging term in the vacuum energy is also absorbed by
microscopic physics to nullify the energy density of
the equilibrium vacuum. The same thermodynamic principle which
leads to the full compensation of the $k_{\rm uv}^4$ term,
leads  to the full compensation of the $k_{\rm uv}^2$ term. Thus
the condensed matter suggests that the thermodynamic principle
of the stability of the free vacuum provides the solution of
both the cosmological constant problem and the hierarchy
problem. 

On the other hand, the logarithmically diverging term
in Eq.(\ref{ZPEnergy2}) is within responsibility of the
effective theory, and it may play an important role when the
deviations from the equilibrium are considered.

\section{Discussion}

There are some  lessons from the condensed matter, and from
quantum hydrodynamics in particular,
for the relativistic quantum fields and gravity. Because of the
power-law divergences, the quantum hydrodynamics cannot be
constructed. One can quantize the acoustic field to obtain its
quanta -- phonons -- and use this only at the tree level. All
other diagrams are not within the responsibility of the
low-energy effective theory. This suggests that the quantum
gravity can only be used  at the tree level too.  The classical
energy functional  (\ref{ClassicalEnergy}) in hydrodynamics, as
well as the classical Einstein action for gravity, and
Eq.(\ref{HiggsHam}) for the Higgs field represent the final
classical output of the whole quantum vacuum.  They cannot be
renormalized by the zero-point energy of the effective quantum
fields,  which present only a part of all the degrees of freedom
of the quantum vacuum. The quadratically and quartically
divergent terms are not within the jurisdiction of the effective
low-energy theory alone. They must
be considered using the microscopic theory of the vacuum state, which is known in condensed matter but is not known in particle physics.

 However, the condensed matter suggests, that the whole quantum
vacuum, which contains the zero-point motion of the effective
quantum fields as well as the trans-Planckian degrees of
freedom, obeys the thermodynamic laws. The latter state that the
energy of the equilibrium free vacuum must be zero. This means
that the huge energy of the zero-point motion of the effective
fields is cancelled without any fine tuning by the
trans-Planckian degrees of freedom. The exact cancellation of
the quartic terms $k_{\rm uv}^4$ in the vacuum energy provides
the possible solution of the cosmological constant problem.
Thus, when the whole vacuum is considered, the natural value of
the cosmological constant $\Lambda$ is zero in the free
equilibrium vacuum. In the perturbed vacuum, $\Lambda$  is
determined not by the quantum zero-point energy, but by the
classical macroscopic parameters of the Universe dynamics:  the
Hubble radius $R$ of the Universe, the Newton constant $G$ and
the energy density of matter $\rho_{\rm mat}$.   
 
Similar exact cancellation of the quadratic terms $k_{\rm
uv}^2$, which are responsible for the radiative corrections to
the Higgs mass, provides the possible solution of the hierarchy
problem in the Standard Model. The superconductivity in metals
and superfluidity in Fermi liquids demonstrate, that  the analog
of the Higgs mass  in these systems is typically much smaller
than the relevant ultraviolet energy scale.

The main condensed-matter argument against the quantum gravity is
that in condensed matter, the effective
metric field is the low energy phenomenon, which naturally emerges  
 together with gauge fields and chiral fermions in
the low-energy corner \cite{Book}. At high energy the metric modes can no
longer be separated from all other microscopic degrees of
freedom of the quantum vacuum, and thus the pure quantum gravity cannot exist. 
 Nevertheless there were attempts
to construct the quantum gravity in terms of the metric field
only, see e.g. Ref. \cite{Litim} where the ultraviolet fixed
point for quantum gravity has been derived. This makes sense
under the following conditions: the ultraviolet fixed point
must occur much below the real microscopic energy scale, i.e.
in the region where the metric field is well
determined; in the infrared limit the cosmological constant
must be zero to match the thermodynamic requirement. One may look for the
similar ultraviolet fixed point in quantum hydrodynamics, i.e. one can try to construct such a liquid in which the
fixed point occurs at the intermediate length scale much bigger
than the interatomic distance, where the macroscopic hydrodynamic
description is still valid.

The possibility of such intermediate energy cut-off scale in
the Standard Model has been discussed in Ref. \cite{Merging}. It
was suggested that the intermediate scale is close to the Planck
scale, and the Newton constant is determined by this scale.
In this scenario, the real microscopic ("atomic") energy scale where, say,
the Lorentz invariance is violated, is much bigger. This scenario
has many advantages. In particular the merging of the running
couplings constants of the weak, strong, and electromagnetic
fields occurs naturally and does not require the unification of
these gauge fields at high energy.

This work is supported in part by the Russian Ministry of
Education and Science, through the Leading Scientific School
grant $\#$2338.2003.2, and by the European Science Foundation 
COSLAB Program.

\end{document}